**Title:** Comparing DWI image quality of deep-learning-reconstructed EPI with RESOLVE in breast lesions at 3.0T: a pilot study

**Authors:** Marialena I. Tsarouchi[1,2], Antonio Portaluri[2,3], Marnix Maas[1], Ritse M. Mann[1,2]

**Affiliations:** [1]Department of Radiology, Nuclear Medicine and Anatomy, Radboud University Medical Center, Nijmegen, the Netherlands; [2]Department of Radiology, the Netherlands Cancer Institute, Amsterdam, the Netherlands; [3] Department of Biomedical Sciences and Morphological and Functional Imaging, University of Messina, Messina, Italy

Synopsis: 100 words (combined 4 sections)

**Motivation:** DWI's challenging spatial resolution could be addressed by deep-learning-based image reconstruction, by reducing noise without increasing acquisition time.

**Goal(s):** To compare the image quality of the Echo-Planar-Imaging-Deep-Learning (EPI-DL) DWI sequence with the clinically used simultaneous-multi-slice (SMS) RESOLVE in breast lesions.

**Approach:** EPI-DL and RESOLVE breast images from 20 participants were qualitatively evaluated ed. Quantitative image quality metrics of SNR and CNR on both high b-value (b800) images and ADC maps were calculated.

**Results:** SNR in RESOLVE vs. EP-DL differed statistically significantly in manually delineations for b800 (p=0.006), ADC maps (p=0.001), and in ADC circularly delineations (p=0.001).

Impact: 40 words

**Impact:** DWI-DL reconstruction may be clinically useful for addressing low-spatial resolution without compromising acquisition time and image quality. Such benefits coupled with the available methods of readout segmentation and SMS acquisitions may further enhance the value of DWI in breast imaging.

Body of the Abstract: 750 words (references not included)

**INTRODUCTION:**

Despite the proven clinical value of Diffusion Weighted Imaging (DWI) in breast imaging, it is still used complementarily to Dynamic Contrast Enhanced (DCE) due to challenging issues (low spatial resolution and poor image quality) that limit DWI's sensitivity [1-3]. Current approaches aim to surpass these downsides of breast DWI and achieve near isotropic spatial resolution comparable with that of DCE in an acceptable acquisition time (< 5 min). However, thus far this is not yet feasible. Deep-learning-based image reconstruction may overcome some of these DWI's hurdles, by reducing noise and improving image quality without prolonging the acquisition time [4, 5].

The aim of this pilot study was to clinically compare the image quality of the Echo Planar Imaging-Deep Learning (EPI-DL) DWI sequence with the standard clinically used simultaneous-multi-slice (SMS) RESOLVE.

## METHODS:

This pilot single-institutional study was approved by the local review committee. Twenty patients, providing informed consent, with histopathologically proven breast lesions, scheduled for breast MRI between May to October 2023, were included.

Breast MRI was performed in a 3.0T MRI system (MAGNETOM Vida XQ-Fit; Siemens Healthcare, Erlangen, Germany). Participants were scanned with the standard clinically applied DWI SMS read-out-segmented, multi-shot echo-planar-imaging (rs-EPI) sequence (RESOLVE; Siemens Healthcare, Erlangen, Germany) and additionally with the prototype deep-learning reconstructed EPI (EPI-DL; Siemens Healthcare, Erlangen, Germany). DW-based acquisition parameters are provided in Table 1. Both DW sequences were performed prior to contrast-agent administration, while standard-of-care clinical protocol followed.

**Table 1: DWI acquisition parameters for standard SMS-RESOLVE and EPI-DL sequences**

| Parameters | SMS-RESOLVE | EPI-DL |
|---|---|---|
| No of slices | 28 | 38 |
| Distance factor | 20% | 20% |
| Phase encoding direction | Posterior to Anterior | Posterior to Anterior |
| Voxel (mm$^3$) | 1.5 x 1.5 x 4.0 | 1.5 x 1.5 x 3.0 |
| TR (ms) | 2670 | 8000 |
| TE (ms) | 55 | 86 |
| FOV (mm$^2$) | 340 x 340 | 340 x 340 |
| Fat suppression | SPAIR | SPAIR |
| Acceleration mode | SMS | GRAPPA |
| Acceleration factor | 2 | 2 |
| Read-out segments | 5 | - |
| Read-out partial Fourier acquisition | 5/8 | - |
| Diffusion Mode | 3D scan trace | 3D scan trace |
| Diffusion directions | 3 orthogonal (x, y, z) | 3 orthogonal (x, y, z) |
| b value (s/mm$^2$); averages | 0; 1 | 0; 1 |
| | 800; 3 | 800; 3 |
| Acquisition time (min:s) | 01:43 | 01:41 |

*Abbreviations: DWI: Diffusion Weighted Imaging; EPI-DL: Echo-Planar-Imaging Deep-Learning; FOV: Field-Of-view; GRAPPA: Generalized Autocalibrating Partially Parallel Acquisitions; SMS: Simultaneous Multi-Slice; SPAIR: Spectral Attenuated Inversion Recovery; TE: Echo Time; TR: Repetition Time.*

Images were anonymized, and lesions were localized based on radiological/pathological information.

A radiologist, blinded to sequence type, qualitatively evaluated the high b-value (i.e., b800) and the ADC maps in both DWI sequences, in terms of overall image quality, lesions' visibility and conspicuity, and presence of artifacts. Images were randomly presented to the radiologist to reduce bias.

Quantitative image quality was evaluated based on Region of Interest (ROIs) of lesions to estimate the signal-to-noise ratio (SNR) and contrast-to-noise ratio (CNR) in b800 and ADC maps in both DWI sequences. In ROI determination two approaches were adopted. A breast imaging expert manually delineated ROIs and placed 2D circular ROIs in solid part of the tumors. SNR was calculated twice as the mean signal intensity (SI) divided by its standard deviation in each ROI, both in b800 and ADC maps. For the CNR, a 2D circular ROI was additionally placed in fibroglandular tissue (FGT). CNR was calculated as absolute difference of SI of the lesion and SI of FGT divided by the lesion's SI standard deviation.

Sequences were evaluated in terms of their diagnostic ability to differentiate benign from malignant breast lesions.

For SNR and CNR comparisons, the paired t-test was exploited for normally distributed data, while the Wilcoxon signed rank test for non-normally distributed data (a=0.05) (according to Shapiro-Wilk test). The independent t-test was used to assess the diagnostic ability of DW-sequences in differentiating benign from malignant breast lesions (a=0.05). Analysis was performed in Matlab (MatLabR2019b, MathWorks, Natick, MA) and SPSS 27.0 (IBM-SPSS Statistics, Armonk, NY).

**RESULTS:**

The 20 participants aged 51±11 years, had 20 histopathologically proven breast lesions (11 benign and 9 malignant).

Qualitative evaluation was in favor of RESOLVE, indicating higher overall image quality, lesion visibility and conspicuity in both b800 and ADC maps. More artifacts were found in EPI-DL both in b800 and ADC maps (Figure 1).

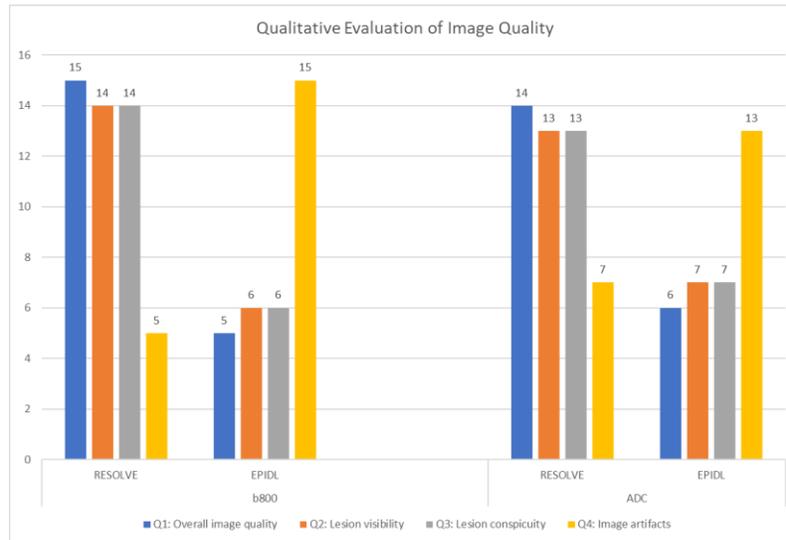

Figure 1: Qualitative evaluation of image quality for b800 and ADC maps in RESOLVE and EPI-DL

The mean lesion size was 65±59 mm$^2$ and 75±67 mm$^2$ for manually and circularly defined ROIs, respectively. Circularly defined FGT regions measured 44±24 mm$^2$. The median SNR defined by manual delineations differed statistically significantly in RESOLVE compared to EP-DL for both b800 (6.2 [IQR=2.2] vs. 4.6 [IQR=2.8]; p=0.006) and ADC maps (5.3 [IQR=4.8] vs. 3.2 [IQR=5.4]; p=0.001). Median SNR defined by circular delineations differed statistically significantly in RESOLVE compared to EPI-DL only for ADC maps (5.4 [IQR=6.2] vs. 3.1 [IQR= 4.8]; p=0.001). CNRs did not reveal any statistically significant difference (p>>0.05).

The mean ADC values in RESOLVE (benign:1.5±0.3 x $10^{-3}$ mm$^2$/s; malignant:1.3±0.4 x$10^{-3}$mm$^2$/s) and in EPI-DL (benign:1.4±0.4 x $10^{-3}$ mm$^2$/s; malignant:1.2±0.4 x$10^{-3}$mm$^2$/s) did not differ statistically significantly (p>>0.05).

**DISCUSSION:** This study evaluated the clinical feasibility of the prototype EPI-DL comparing to the currently utilized SMS-RESOLVE. Although RESOLVE proved to maintain high overall image quality, EPI-DL achieves comparably good image quality, in accordance with the literature [4,5]. It seems advisable to integrate the benefits of DL reconstruction with the available methods to improve DWI image quality including read-out-segmentation and SMS acquisitions to further enhance the value of DWI in breast imaging.

**CONCLUSION:** Breast DWI-DL reconstruction may be clinically useful for addressing low-spatial resolution without compromising acquisition time and image quality.

**REFERENCES:**

1. Iima M, Honda M, Sigmund EE, Ohno Kishimoto A, Kataoka M, Togashi K. Diffusion MRI of the breast: Current status and future directions. J Magn Reson Imaging. 2020;52(1):70-90.

2. Gullo RL, Partridge SC, Shin HJ, Thakur SB, et al. Update on DWI for Breast Cancer Diagnosis and Treatment Monitoring [published online ahead of print]. AJR Am J Roentgenol. 2023; 10.2214/AJR.23.29933.3. Baltzer P, Mann RM, Iima M, et al. Diffusion-weighted imaging of the breast-a consensus and mission statement from the EUSOBI International Breast Diffusion-Weighted Imaging working group. Eur Radiol. 2020;30(3):1436-1450.

4. Wilpert C, Neubauer C, Rau A, et al. Accelerated Diffusion-Weighted Imaging in 3 T Breast MRI Using a Deep Learning Reconstruction Algorithm With Superresolution Processing: A Prospective Comparative Study [published online ahead of print]. Invest Radiol. 2023;10.1097/RLI.0000000000000997.

5. Wessling D, Gassenmaier S, Olthof SC, et al. Novel deep-learning-based diffusion weighted imaging sequence in 1.5 T breast MRI. Eur J Radiol. 2023;166:110948.